\documentclass[aps,prb,reprint,longbibliography,superscriptaddress]{revtex4-2}
\usepackage{amsmath,amssymb,bm}
\usepackage{graphicx}
\usepackage{hyperref}
\usepackage{xcolor}
\usepackage{color}

\begin{document}

\title{Anomalous field evolution of the mixed-state linewidth in the second superconducting dome of LaFeAsO$_{1-x}M_x$ (\mbox{$M={\rm F,H}$})}

\author{Rustem Khasanov}
 \email{rustem.khasanov@psi.ch}
 \affiliation{PSI Center for Neutron and Muon Sciences CNM, 5232 Villigen PSI, Switzerland}

\author{Pierre Dalmas de R\'{e}otier}
 \affiliation{Universit\'{e} Grenoble Alpes, CEA, Grenoble INP, IRIG-PHELIQS, 38000 Grenoble, France}

\author{Samuele Sanna}
\affiliation{Dipartimento di Fisica e Astronomia ``Augusto Righi'',
Universit\`a di Bologna, 40127 Bologna, Italy}

\author{Gianrico Lamura}
\affiliation{CNR-SPIN, Via Dodecaneso 33, 16152 Genova, Italy}

\author{Hubertus Luetkens}
 \affiliation{PSI Center for Neutron and Muon Sciences CNM, 5232 Villigen PSI, Switzerland}


\author{Matteo Moroni}
 \affiliation{Dipartimento di Fisica, Università degli Studi di Pavia, 27100 Pavia, Italy}

\author{Pietro Carretta}
 \affiliation{Dipartimento di Fisica, Università degli Studi di Pavia, 27100 Pavia, Italy}

\author{Rhea Kappenberger}
 \affiliation{Leibniz-Institute for Solid State and Materials Research, IFW-Dresden, 01069 Dresden, Germany}

\author{Rowena Wachtel}
 \affiliation{Leibniz-Institute for Solid State and Materials Research, IFW-Dresden, 01069 Dresden, Germany}

\author{Bernd B\"uchner}
 \affiliation{Leibniz-Institute for Solid State and Materials Research, IFW-Dresden, 01069 Dresden, Germany}

\author{Sabine Wurmehl}
 \affiliation{Leibniz-Institute for Solid State and Materials Research, IFW-Dresden, 01069 Dresden, Germany}

\author{Nikolai D. Zhigadlo}
 \affiliation{CrystMat Company, 8037 Zurich, Switzerland}

\date{\today}

\begin{abstract}
We report a transverse-field muon-spin rotation/relaxation ($\mu$SR) study of the internal-field distribution in the mixed state of LaFeAsO$_{0.89}$F$_{0.11}$ and LaFeAsO$_{0.75}$H$_{0.25}$, representative of the first (SC1) and second (SC2) superconducting domes of the LaFeAsO$_{1-x}M_x$ ($M={\rm F,H}$) family, respectively.
Below the superconducting transition temperature $T_{\rm c}$, the linewidth of the internal-field distribution increases in both samples, indicating the formation of a vortex lattice.
Above $T_{\rm c}$, the linewidth remains field dependent and increases approximately linearly with field, consistent with broadening of the powder spectrum caused by an anisotropic Knight shift.
After subtraction of this normal-state contribution, the superconducting linewidth $\sigma_{\rm sc}$ exhibits qualitatively different field dependences in the two samples.
At 4~K, the SC1 ($x_{\rm F}=0.11$) sample shows the expected monotonic decrease with increasing field, whereas the SC2 ($x_{\rm H}=0.25$) sample develops a pronounced local maximum near 3~T.
A contour representation of $\sigma_{\rm sc}(T,H)$ further reveals a ridge of local maxima whose field position, $H_{\sigma,\max}(T)$, shifts to lower fields upon warming and disappears near $T_{\rm c}$.
The anomalous field evolution observed in the SC2 sample is consistent with an additional field-induced contribution associated with enhanced Pauli-paramagnetic effects, highlighting the distinct electronic character of the two superconducting domes.
\end{abstract}
\maketitle

\begin{figure}[htb]
\includegraphics[width=1.0\linewidth]{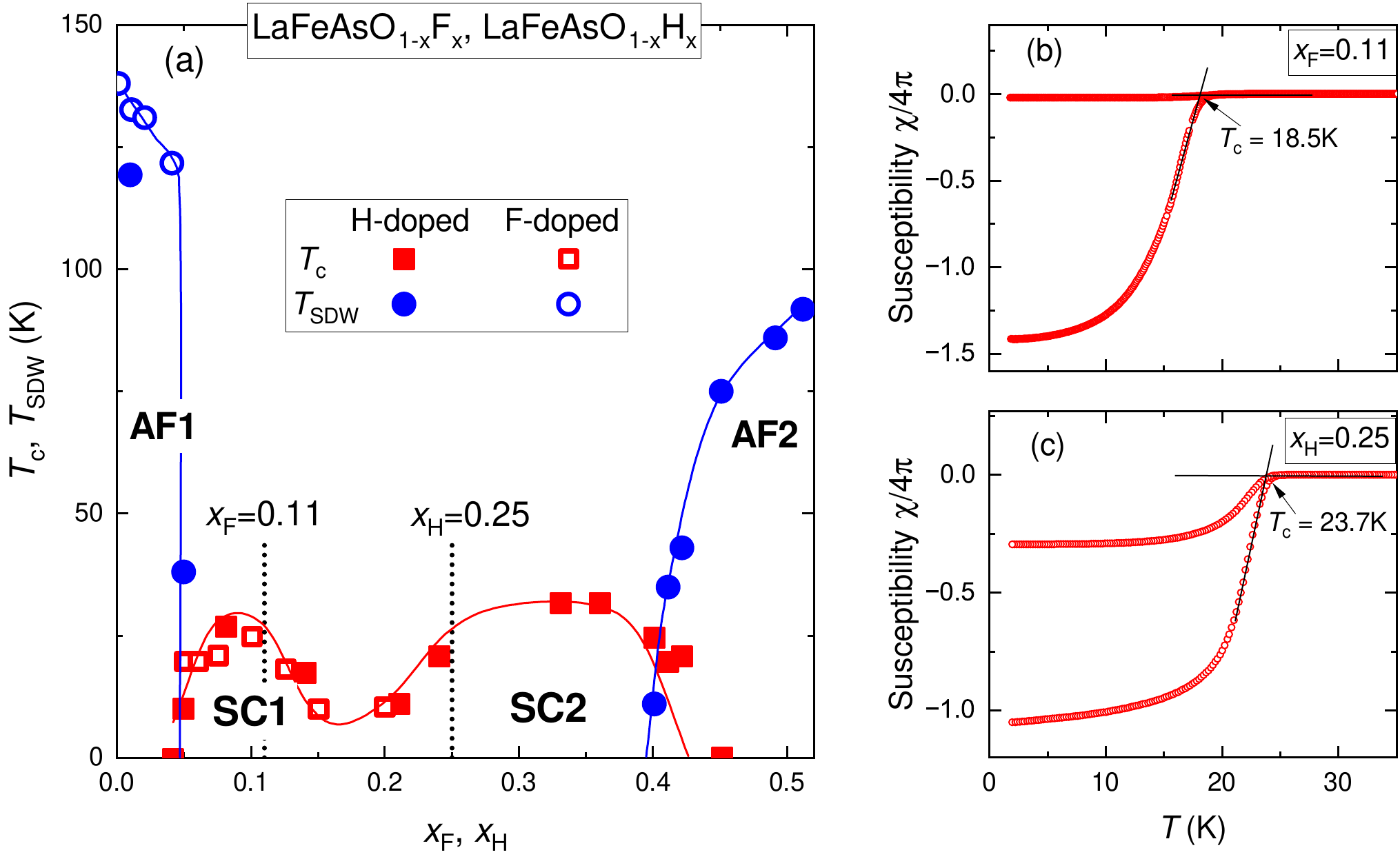}
\caption{(a) Schematic phase diagram of LaFeAsO$_{1-x}M_x$ ($M={\rm F,H}$), based on Refs.~\onlinecite{Luetkens2009, Lamura2014, Iimura2012, HosonoKuroki2015, Hiraishi2014}. The dotted vertical lines mark the two compositions studied in the present work: $x_{\rm F}=0.11$ (LaFeAsO$_{0.89}$F$_{0.11}$), located in the first superconducting dome SC1, and $x_{\rm H}=0.25$ (LaFeAsO$_{0.75}$H$_{0.25}$), located in the second superconducting dome SC2. (b) and (c) Zero-field-cooled (ZFC) and field-cooled (FC) magnetic susceptibility curves for the $x_{\rm F}=0.11$ and $x_{\rm H}=0.25$ samples, measured in applied fields of 2 and 0.3~mT, respectively. The straight lines illustrate the linear extrapolation procedure used to determine the superconducting transition temperatures $T_c=18.5$ and 23.7~K.}
\label{fig1}
\end{figure}

\noindent \underline{{\it Introduction.}} The LaFeAsO$_{1-x}M_x$ ($M={\rm F,H}$) iron pnictides, commonly referred to as the La1111 family, remain a central platform for investigating the interplay between superconductivity, magnetism, multiorbital electronic structure, and electronic correlations in Fe-based superconductors \cite{Kamihara2008,HosonoKuroki2015}. In the fluorine-substituted series LaFeAsO$_{1-x}$F$_x$, electron doping suppresses the antiferromagnetic phase of the parent compound and gives rise to superconductivity in the conventional low-doping regime~\cite{Kamihara2008,Luetkens2009}. Hydrogen substitution  in LaFeAsO$_{1-x}$H$_x$ extends the accessible electron-doping range much further and reveals a richer phase diagram, where superconductivity separates into two domes, SC1 and SC2, and is flanked by two antiferromagnetic regions, AF1 and AF2, see Refs.~\onlinecite{Iimura2012, HosonoKuroki2015, Hiraishi2014} and Fig.~\ref{fig1}. An important point is that, up to $x\simeq0.2$, F and H substitution lead essentially to the same phase diagram, with hydrogen acting as an indirect electron dopant analogous to fluorine but with a much larger solubility range~\cite{HosonoMatsuishi2013,Iimura2012, Lamura2014}. Only at higher doping does the H-substituted system reveal the second superconducting dome, SC2, and the neighboring antiferromagnetic phase AF2~\cite{Hiraishi2014}.

The existence of the two superconducting domes indicates that superconductivity in La1111 evolves in qualitatively different electronic environments \cite{Iimura2012,HosonoKuroki2015}. SC1 corresponds to the low-doping regime that is closely connected to the widely studied LaFeAsO$_{1-x}$F$_x$ phase diagram, whereas SC2 develops at substantially higher electron doping \cite{Iimura2012, HosonoMatsuishi2013}.
Previous experimental and theoretical studies have shown that the high-doping regime is characterized by weakened hole-electron nesting, a more incoherent normal state, and enhanced correlation effects associated with the Fe $3d_{xy}$ band~\cite{Iimura2012,Iimura2016,HosonoKuroki2015}.
Moreover, SC2 is adjacent to the second antiferromagnetic phase AF2, whose microscopic origin appears distinct from that of the low-doping AF1 state and is argued to be more strongly influenced by electronic correlations~\cite{Hiraishi2014,Fujiwara2013,Hiraishi2020}. These observations suggest that SC2 is not merely a continuation of SC1, but rather a superconducting state developing in a different region of the phase diagram, with a modified balance between itinerancy, correlations, and magnetism~\cite{HosonoKuroki2015, Hiraishi2020}.

The distinction between SC1 and SC2 is also reflected in their response to magnetic field. Recent high-field measurements on LaFeAsO$_{1-x}$H$_x$ have shown that the dominant pair-breaking mechanism differs markedly between the two domes: while the low-doping SC1 regime is governed primarily by the orbital effect, the SC2 regime exhibits a much stronger paramagnetic contribution~\cite{Kawachi2023}. This difference is particularly suggestive in the context of transverse-field (TF) muon-spin rotation/relaxation ($\mu$SR), which directly probes the internal-field distribution in the mixed state~\cite{Sonier2000,Brandt1988}. For a conventional type-II superconductor, the superconducting contribution to the mixed-state linewidth is expected to evolve smoothly and monotonically with field~\cite{Brandt1988,Brandt2003}.
By contrast, in superconductors with strong Pauli-paramagnetic effects, the Zeeman currents can substantially modify the field distribution around vortices and generate a local maximum in the linewidth~\cite{Ichioka2007, Dalmas2011, MichalMineev2010, Bianchi2008, Spehling2009, White2010}. These considerations naturally raise the question of whether the mixed-state response remains conventional in both superconducting domes of La1111, or whether the SC2 regime exhibits an additional field-induced contribution absent in SC1.

To address this issue, we compare the mixed-state linewidth in representative SC1 (LaFeAsO$_{0.89}$F$_{0.11}$, $x_{\rm F}=0.11$) and SC2 (LaFeAsO$_{0.75}$H$_{0.25}$, $x_{\rm H}=0.25$) compositions of La1111, marked by the dotted lines in Fig.~1(a). Using TF-$\mu$SR, we show that the SC1 ($x_{\rm F}=0.11$) sample exhibits the expected conventional monotonic field evolution of the superconducting linewidth, whereas the SC2 ($x_{\rm H}=0.25$) sample develops a pronounced local maximum. We further demonstrate that the field at which this maximum occurs shifts systematically to lower values with increasing temperature. We argue that this behavior is a hallmark of the mixed state in the SC2 phase and reflects an additional field-induced contribution to the internal-field distribution that is absent in the SC1 phase.

\noindent \underline{\it Experimental details.}
The transverse-field (TF) muon-spin rotation/relaxation ($\mu$SR) experiments were performed on the High-Field and Low-Temperature spectrometer HAL-9500 at the Swiss Muon Source (S$\mu$S), Paul Scherrer Institute (PSI), Villigen, Switzerland~\cite{Stoykov2012,HAL9500}. The $\mu$SR data were analyzed using the \textsc{MUSRFIT} software package~\cite{MUSRFIT}. The experimental setup and the data-analysis procedure are described in the Supplemental Information.

The samples were prepared following the procedures described in Refs.~\onlinecite{Khasanov2011,Zhigadlo2012,Shiroka2018,Qureshi2010,Alfonsov2011}; further details are provided in the Supplemental Information.

The dc magnetization measurements were performed using a commercial SQUID magnetometer. The zero-field-cooled (ZFC) and field-cooled (FC) magnetic susceptibility curves are shown in Figs.~\ref{fig1}(b) and \ref{fig1}(c). The measurements were carried out in applied magnetic fields of 2 and 0.3~mT for the $x_{\rm F}=0.11$ and $x_{\rm H}=0.25$ samples, respectively. Both samples exhibit bulk superconductivity, as evidenced by the low-temperature susceptibility approaching full diamagnetic screening. The superconducting transition temperatures, $T_{\rm c}$, determined from linear fits to the susceptibility data in the vicinity of the transition, were estimated to be approximately 18.5 and 23.7~K for the $x_{\rm F}=0.11$ and $x_{\rm H}=0.25$ samples, respectively.

\begin{figure*}[htb]
\includegraphics[width=1.0\linewidth]{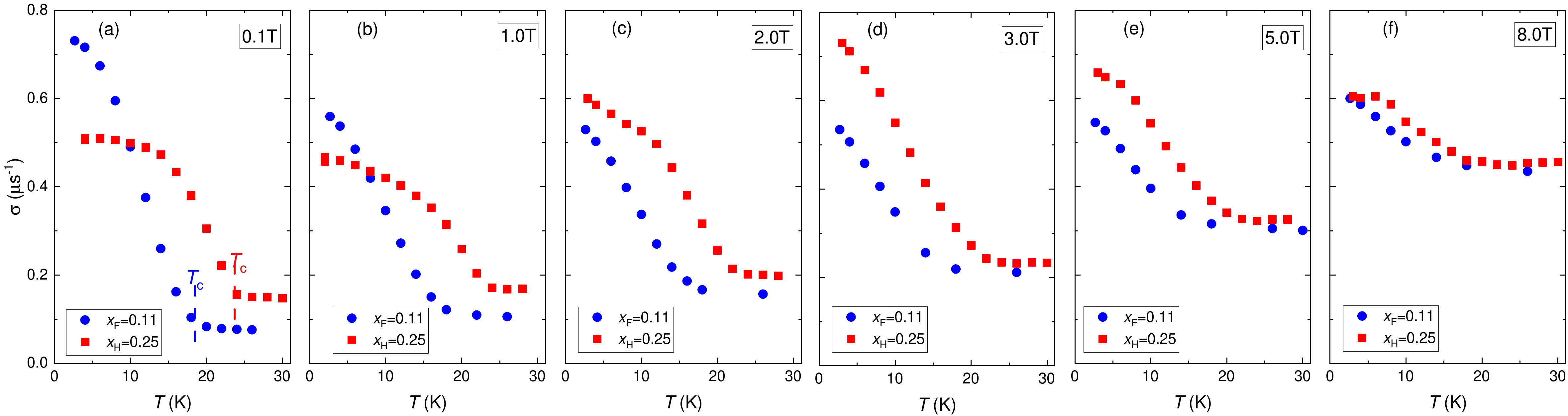}
\caption{(a) to (f) Temperature dependence of the Gaussian relaxation rate $\sigma(T)$ measured in transverse field for $x_{\rm F}=0.11$ (LaFeAsO$_{0.89}$F$_{0.11}$) and $x_{\rm H}=0.25$ (LaFeAsO$_{0.75}$H$_{0.25}$) at representative fields between 0.1 and 8.0~T.  Dotted lines in (a) indicate the approximate position of $T_{\rm c}$ as obtained in magnetization experiments.}
\label{fig2}
\end{figure*}

\noindent \underline{\it Results and Discussion.} In the present experiments the mixed-state linewidth is tracked through the Gaussian relaxation rate $\sigma$ extracted from the TF-$\mu$SR spectra. Figure~\ref{fig2} summarizes the temperature dependence of $\sigma$ measured at several magnetic fields between 0.1 and 8~T for the two samples. In both compounds, $\sigma(T)$ begins to increase below $T_c$, consistent with the onset of the vortex-state field inhomogeneity \cite{Pumpin1990, Sonier2000, Brandt1988, Brandt2003, Blundell_book_2022, Amato-Morenzoni_book_2024}. Above $T_c$, $\sigma(T)$ tends to a plateau at each field, which would ordinarily be associated with the quasistatic nuclear contribution. In the present case, however, the normal-state linewidth itself is field dependent, thus indicating that, in addition to the nuclear contribution, the polycrystalline spectra contain a field-dependent broadening.

\begin{figure*}[htb]
\includegraphics[width=0.85\linewidth]{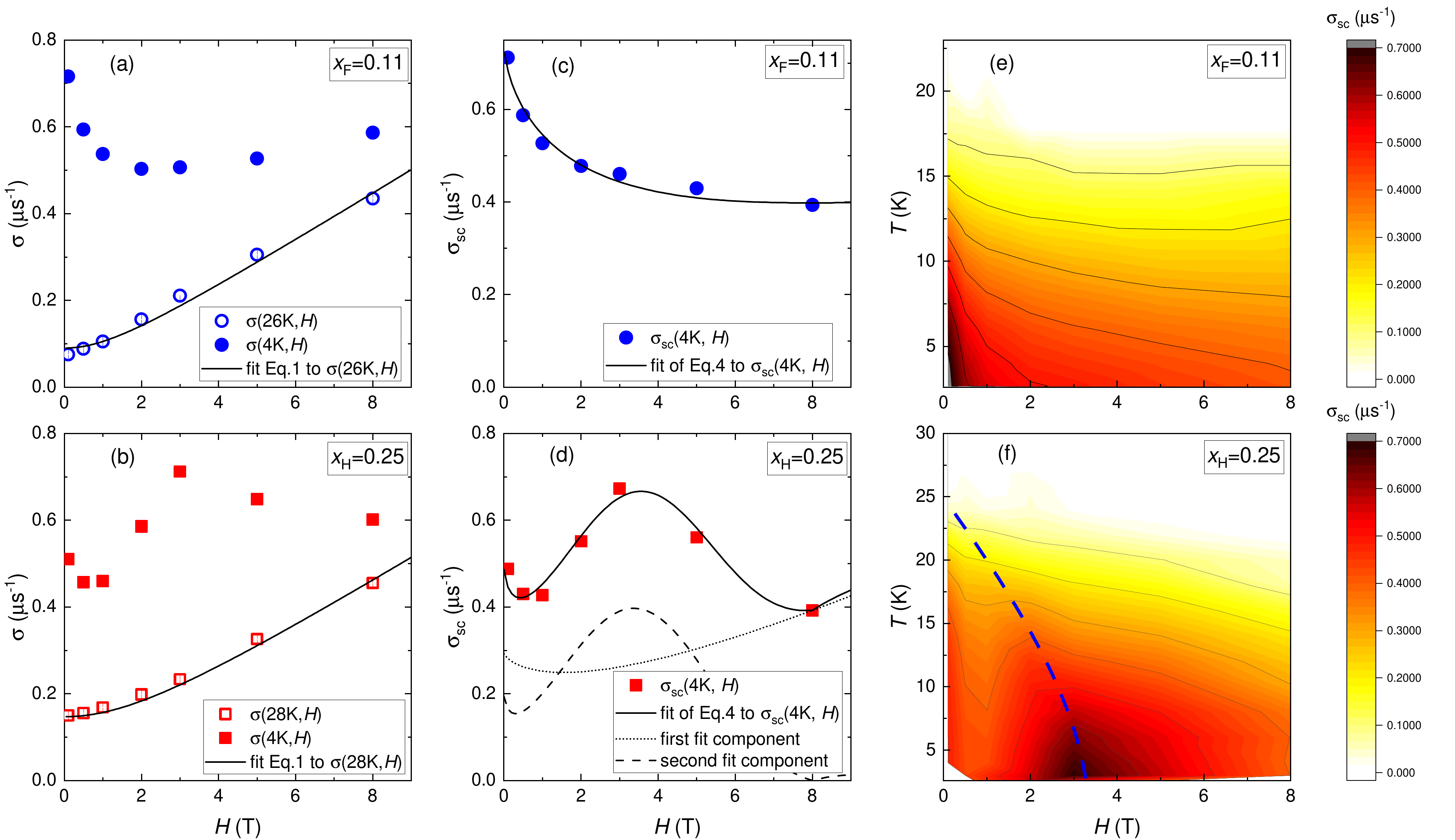}
\caption{(a) and (b) Field dependences of the Gaussian relaxation rate $\sigma$ measured above $T_c$ ($T=26$~K for $x_{\rm F}=0.11$ and $T=28$~K for $x_{\rm H}=0.25$) together with the corresponding data taken below $T_{\rm c}$ at $T=4$~K. The solid lines are fits of Eq.~(\ref{eq:normal}) to the normal-state data.
(c) and (d) The superconducting contribution to the $\mu$SR linewidth, $\sigma_{\rm sc}$, at $T=4$~K, obtained by subtracting the normal-state contribution in quadrature according to Eq.~(\ref{eq:sigmasc}). The solid lines are fits based on the model of Ref.~\onlinecite{Dalmas2011}; see text for further details.
(e) and (f)  Contour plots of $\sigma_{\rm sc}(T,H)$ for $x_{\rm F}=0.11$ and $x_{\rm H}=0.25$, respectively. For $x_{\rm H}=0.25$, the dashed line traces the field position $H_{\sigma,\max}(T)$ of the ridge of local maxima.}
\label{fig3}
\end{figure*}

To emphasize the different low- and high-temperature behavior, Figs.~\ref{fig3}(a) and \ref{fig3}(b) show the field dependences of $\sigma$ measured above $T_{\rm c}$ ($T=26$~K for the $x_{\rm F}=0.11$ sample and $T=28$~K for the $x_{\rm H}=0.25$ sample), together with the corresponding data taken below $T_{\rm c}$ at $T=4$~K. To account for the field dependence in the normal state, the data above $T_c$ were fitted by
\begin{equation}
\sigma_{\rm n}(H)=\sqrt{\sigma_{\rm nm}^2+(kH)^2},
\label{eq:normal}
\end{equation}
where $\sigma_{\rm n}$ is the normal-state contribution, $\sigma_{\rm nm}$ is the field-independent nuclear contribution, and $k$ describes the linear broadening associated with the anisotropic Knight shift. As shown by the solid lines in Figs.~\ref{fig3}(a) and \ref{fig3}(b), Eq.~(\ref{eq:normal}) describes the normal-state data of both samples reasonably well. Although the two compounds show a modest difference in $\sigma_{\rm nm}$, namely $0.10(1)~\mu{\rm s}^{-1}$ for $x_{\rm F}=0.11$ and $0.15(1)~\mu{\rm s}^{-1}$ for $x_{\rm H}=0.25$, the coefficient $k=0.0548(4)~\mu{\rm s}^{-1}{\rm T}^{-1}$ is the same within the experimental uncertainty, indicating that the field-dependent normal-state broadening mechanism is essentially identical in the two powder samples. By contrast, the low-temperature behavior is markedly different: for $x_{\rm F}=0.11$, $\sigma(4{\rm K},H)$ decreases rapidly with increasing field and then shows a slight upturn above a few tesla, whereas for $x_{\rm H}=0.25$ it decreases only weakly at low fields, then increases, reaches a pronounced maximum near 3~T, and decreases again at higher fields.

The superconducting contribution to the $\mu$SR linewidth was then obtained by subtracting the normal-state contribution in quadrature \cite{Pumpin1990, Yaouanc1997, YaouancBook2011, Blundell_book_2022, Amato-Morenzoni_book_2024},
\begin{equation}
\sigma_{\rm sc}(T,H)=\sqrt{\sigma^2(T,H)-\sigma_{\rm n}^2(H)}.
\label{eq:sigmasc}
\end{equation}
The resulting low-temperature field dependences of $\sigma_{\rm sc}(4{\rm K}, H)$ are shown in Figs.~\ref{fig3}(c) and \ref{fig3}(d). The $x_{\rm F}=0.11$ sample displays the expected monotonic decrease of $\sigma_{\rm sc}$ with increasing field \cite{Brandt1988,Brandt2003}. In contrast, the $x_{\rm H}=0.25$ sample exhibits a pronounced local maximum centered near 3~T.

Because a nonmonotonic field dependence of $\sigma_{\rm sc}(H)$ is precisely the behavior expected when Zeeman-related currents become important in a superconductor with strong Pauli-paramagnetic effects~\cite{Ichioka2007,Dalmas2011,MichalMineev2010,Bianchi2008,Spehling2009,White2010}, we analyze the data in Figs.~\ref{fig3}(c) and \ref{fig3}(d) within the framework developed by Dalmas de R\'eotier and Yaouanc~\cite{Dalmas2011}. In that work, the field distribution in a hard type-II superconductor is described within an approximate Ginzburg-Landau treatment that explicitly includes both the conventional orbital supercurrents and an additional Zeeman-current contribution~\cite{MichalMineev2010}. The theory is formulated in terms of the Fourier components $B_{\mathbf K}$ of the vortex-lattice field distribution, written as the sum of orbital and Zeeman terms,
\begin{equation}
B_{\mathbf K}=B_{{\mathbf K},{\rm orb}}+B_{{\mathbf K},{\rm Zee}}.
\end{equation}
The normalized form factor is then controlled by the reduced field $h=H/H_{c2}$ and by the dimensionless parameter $R=C_{\rm Zee}/B_L$, where $C_{\rm Zee}$ is the coefficient of the Zeeman term and $B_L$ sets the orbital field scale~\cite{Dalmas2011,MichalMineev2010}. A key result of the theory is that, when $R$ becomes sufficiently large, the linewidth of the field distribution may develop a maximum at an intermediate reduced field instead of decreasing monotonically with increasing field. Since the TF-$\mu$SR linewidth is directly related to the second moment of the internal-field distribution~\cite{YaouancBook2011}, this framework provides a natural basis for analyzing the anomalous $\sigma_{\rm sc}(H)$ behavior observed here. We note, however, that the original model is formally justified near $T_c$ and for $T>T^\ast \simeq 0.56\,T_c$~\cite{Dalmas2011}; therefore, in the present case it should be regarded primarily as a phenomenological description of the field evolution rather than as a fully quantitative microscopic treatment.

In analogy with the phenomenological $\alpha$-model commonly used for multigap superconductors~\cite{Bouquet2001, Carrington2003, Khasanov2007, Khasanov2009}, we parameterized the normalized field dependence of the superconducting linewidth at 4~K as a weighted sum of two contributions,
\begin{equation}
\frac{\sigma_{\rm sc}(H)}{\sigma_{\rm sc}(0)}
=
\omega\,
\frac{\sigma_{{\rm sc},1}(H,H_{c2,1},R_1)}
{\sigma_{{\rm sc},1}(0,H_{c2,1},R_1)}
+
(1-\omega)\,
\frac{\sigma_{{\rm sc},2}(H,H_{c2,2},R_2)}
{\sigma_{{\rm sc},2}(0,H_{c2,2},R_2)},
\label{eq:two_comp_sigma}
\end{equation}
where $\sigma_{{\rm sc},i}(H,H_{c2,i},R_i)$ denotes the single-component field dependence calculated within the scheme of Ref.~\onlinecite{Dalmas2011}, $H_{c2,i}$ is the upper critical field, $R_i$ quantifies the relative weight of the Zeeman contribution, and $\omega$ is the relative weight of the first component.

To reduce the number of adjustable parameters, the upper critical field scales were not treated as free parameters but were fixed to representative experimental values, namely $H_{c2}=60$~T for LaFeAsO$_{0.89}$F$_{0.11}$~\cite{Hunte2008,Kohama2009} and $H_{c2}=75$~T for LaFeAsO$_{0.75}$H$_{0.25}$~\cite{Kawachi2023}. For the $x_{\rm F}=0.11$ sample, the data are well described by a single component with $H_{c2,1}=60$~T, $R_1=19.0$, and $\omega=1.0$. By contrast, the $x_{\rm H}=0.25$ sample requires two components with $H_{c2,1}=75$~T, $R_1=220$, $H_{c2,2}=8$~T, $R_2=120$, and $\omega=0.60$. The corresponding fits and the individual fit components are shown in Figs.~\ref{fig3}(c) and \ref{fig3}(d) by solid, dotted, and dashed lines. While the detailed microscopic interpretation of this two-component description remains open, the failure of a single-component description for the $x_{\rm H}=0.25$ sample clearly shows that its field distribution cannot be understood as a simple extension of the more conventional SC1 behavior. The substantially larger fitted $R$ values obtained for the $x_{\rm H}=0.25$ sample further indicate that the anomalous field evolution in SC2 is associated with a much stronger Zeeman contribution to the mixed-state field distribution.

The temperature evolution of $\sigma_{\rm sc}(T,H)$ is illustrated more clearly by the contour plots in Figs.~\ref{fig3}(e) and \ref{fig3}(f). For the $x_{\rm H}=0.25$ sample, the local maximum in $\sigma_{\rm sc}(H)$ is not fixed at a single field, but instead defines a temperature-dependent position $H_{\rm \sigma,max}(T)$ that shifts to lower field on warming. A simple interpretation is that the local maximum occurs at a roughly fixed reduced field,
\begin{equation}
H_{\rm \sigma,max}(T)\simeq b^{\ast}H_{c2}(T), \nonumber
\end{equation}
with $b^{\ast}$ being a characteristic fraction of the upper critical field. Since $H_{c2}(T)$ decreases with increasing temperature, $H_{\rm \sigma,max}(T)$ is expected to move to lower values, exactly as observed. In this picture, the local maximum in $\sigma_{\rm sc}(T,H)$ marks the field at which the Zeeman-related contribution becomes the dominant term relative to the orbital vortex contribution.

A closely related interpretation is to regard $H_{\rm \sigma,max}(T)$ as the balance point between two competing contributions. At low fields, the linewidth is governed mainly by the ordinary vortex-lattice response. As the field increases, the Zeeman-related contribution grows and eventually becomes strong enough to reverse the trend. At still higher fields, vortex-core overlap and the overall reduction of the superconducting response suppress the linewidth again. Upon warming, the orbital contribution weakens, and this balance is reached at lower field, leading naturally to the observed shift of $H_{\rm \sigma,max}$.

These interpretations are physically plausible in view of the distinct electronic environments of SC1 and SC2. As discussed above, SC2 develops in a more strongly correlated and electronically less coherent regime, with weakened hole-electron nesting~\cite{Iimura2012,Iimura2016,Hiraishi2020}. Moreover, recent high-field measurements have shown that the paramagnetic contribution to pair breaking is much stronger in SC2 than in SC1~\cite{Kawachi2023}. These characteristics make SC2 a natural setting for a sizable Zeeman-related modification of the mixed-state field distribution, whereas SC1 remains dominated by more conventional orbital physics.

\noindent \underline{\it Conclusions.}
We have investigated the field evolution of the mixed-state linewidth in two representative compositions taken from the first and second superconducting domes of LaFeAsO$_{1-x}M_x$ ($M={\rm F,H}$), namely LaFeAsO$_{0.89}$F$_{0.11}$ (SC1) and LaFeAsO$_{0.75}$H$_{0.25}$ (SC2). In both samples, the normal-state linewidth is described by a field-independent nuclear contribution together with a linear field-dependent broadening attributed to an anisotropic Knight shift. After subtraction of this normal-state contribution in quadrature, the superconducting contribution to the linewidth, $\sigma_{\rm sc}$, shows qualitatively different field dependences in the two systems.
For the $x_{\rm F}=0.11$ sample, $\sigma_{\rm sc}(H)$ follows the conventional monotonic behavior expected for a vortex-lattice field distribution dominated by the orbital response. By contrast, the $x_{\rm H}=0.25$ sample exhibits a pronounced nonmonotonic field dependence, with a local maximum occurring near $3$~T at $4$~K. Within the phenomenological analysis based on the approach of Dalmas de R\'eotier and Yaouanc \cite{Dalmas2011}, the $x_{\rm F}=0.11$ data are well described by a single component, whereas the $x_{\rm H}=0.25$ data require a two-component description with substantially larger values of the Zeeman-related parameter $R$.

The contour representation of $\sigma_{\rm sc}(T,H)$ provides further evidence that the anomaly in the $x_{\rm H}=0.25$ sample is intrinsic to the superconducting state: the field position of the local maximum defines a temperature-dependent characteristic field $H_{\rm \sigma,max}(T)$ that shifts systematically to lower values upon warming.  Taken together, these results show that the mixed-state response in SC2 differs qualitatively from that in SC1. The most natural interpretation is that, in addition to the ordinary vortex-lattice contribution, the SC2 state contains an extra field-induced term associated with a strongly enhanced Zeeman response.

\noindent \underline{\it Acknowledgments.} This work was performed at the Swiss Muon Source S$\mu$S, Paul Scherrer Institute, Villigen, Switzerland. 
The Dresden group thanks D.~Meiler, C.~G.~F.~Blum, L.~Giebeler, S.~M\"uller-Litvanyi, G.~Kreutzer, and S.~Gass for technical support at IFW Dresden. The work in Dresden was supported by the Deutsche Forschungsgemeinschaft (DFG) through the Priority Programme SPP~1458 under Grants No.~BE~1749/13-1 and BU~887/15-1, through SFB~1143 (Project-ID 247310070, project B01), through the Research Training Group GRK~1621, through the Emmy Noether Programme under Grant No.~WU~595/3-1, and through DFG Grant No.~WU~595/17-1 (Project-ID 461247437).

\end{document}